\documentclass[a4paper,12pt]{article}
\usepackage[left=2cm,right=2cm]{geometry}
\usepackage{graphics}
\newcommand{\fpi}{f_\pi^2}
\newcommand{\fk}{f_K^2}
\newcommand{\mpi}{m_\pi^2}
\newcommand{\mk}{m_K^2}
\newcommand{\sq}{\mpi + \mk}
\newcommand{\nn}{\nonumber}
\title{The $\sigma K$ coupling in the chiral unitary approach and the isoscalar $\bar{K}N$, $\bar{K}A$ interaction}
\author{A. Mart\'inez Torres, K. P. Khemchandani and E. Oset}

\date{}

\begin{document}
\maketitle
\begin{center}
Departamento de F\'{\i}sica Te\'orica and IFIC,
Centro Mixto Universidad de Valencia-CSIC, Institutos de
Investigaci\'on de Paterna, Aptdo. 22085, 46071 Valencia, Spain
\end{center}
\abstract{
We evaluate the ``$\sigma$" exchange contribution to the $\bar{K}N\rightarrow\bar{K}N$ scattering within a chiral unitary approach. We show that the chiral transition potentials for $\pi \pi \rightarrow K \bar{K}$ in the $t$-channel lead to a ``$\sigma$" contribution that vanishes in the $\bar{K}$ forward direction and, hence, would produce a null ``$\sigma$" exchange contribution to the $K^-$ optical potential in nuclear matter in a simple impulse approximation. This is a consequence of the fact that the leading order chiral Lagrangian gives an $I=0$ $\pi\pi\rightarrow K\bar{K}$ amplitude proportional to the squared momentum transfer, $q^2$. This finding poses questions on the meaning or the origin of ``$\sigma$" exchange potentials used in relativistic mean field approaches to the $K^-$ nuclear selfenergy. This elementary ``$\sigma$" exchange potential in $\bar{K}N\rightarrow\bar{K}N$ is compared to the Weinberg-Tomozawa term and is found to be smaller than present theoretical uncertainties but will be relevant in the future when aiming at fitting increasingly more accurate data.
} 

\section{Introduction}
A lot of attention is being paid to the interaction of the 
negative kaons with nucleons nowadays. 
It has been traditionally a testing field for chiral dynamics and
 non-perturbative methods since, close below the threshold, there is the $\Lambda(1405)$ resonance, 
 which is dynamically generated within the non-perturbative chiral unitary approach
  \cite{weise,angels,Oller:2000fj,Oset:2001cn}. 
 Much before these methods became popular, the $\Lambda(1405)$ was already advocated as a bound state of meson
 baryon in the coupled channels of $\pi \Sigma $ and $\bar{K}N$
  \cite{dalitz,jennings}. One of the consequences of the systematic study of the
 $\bar{K} N$ interaction in the chiral unitary approach is the realization that
 there are two poles for the  $\Lambda(1405)$ resonance 
 \cite{jido,Garcia-Recio:2002td,Garcia-Recio:2003ks,Hyodo:2002pk,GarciaRecio:2005hy,Hyodo:2007jq},
which get support from the recent experiment \cite{prakhov} in
conjunction with the analysis done in \cite{magas}.
 The recent determination of the $K^- p$ scattering 
 length from  the study of $K^- p$ atoms in DEAR at DAFNE \cite{Beer:2005qi} has stimulated 
 a revival of the work
 on this issue  and several works have been done 
  \cite{Borasoy:2005ie,Oller:2005ig,Oller:2006jw,borasoy} by
 including chiral Lagrangians
 of higher order to the lowest order one used in \cite{angels,Oset:2001cn}.
 
  Simultaneously, much work has also been done  along these lines in order to
study the interaction of kaons with nuclei, producing a $K^-$ nucleus optical
potential, which is currently used to study kaonic atoms and to  make prospects for
the possible kaon-nuclear deeply bound states. The so called chiral potentials 
 \cite{lutz,angelsself,schaffner,galself}  lead to
a moderate attraction of about 50 MeV at normal nuclear matter density and differ
appreciably from the $t \rho$ approximation, since at the threshold 
the $\bar{K}N$ $t$-matrix is repulsive.  The selfconsistency of the
calculation is an important requirement for the construction of the potential
due to the presence of the $\Lambda(1405)$ resonance below the threshold, and is
responsible for a fast transition from the repulsive potential in 
the $t \rho$ approximation at very low
densities to an attraction at the densities felt by measured  kaonic atom
states. With this``shallow" theoretical potential a good reproduction of the data
on shifts and widths of kaonic atoms was reported in \cite{okumura}. On the other
hand, in \cite{baca}, starting from the theoretical potential of
 \cite{angelsself}, a small phenomenological part was added to that potential
and a fit was conducted to the full set of kaonic atoms data, concluding that
corrections needed for such a best fit were of the order of 20 \% of the theoretical potential. With this
potential, plus the calculated imaginary part of about (-)50 MeV at normal  nuclear
matter, one obtains also deeply bound  $K^-$ nuclear states, but with a width much
bigger than the binding energy ($\Gamma \sim -2 Im\{V_{opt}\}$), which would 
preclude the experimental observation of the corresponding peaks. 

    Parallelly, some highly attractive potential with about
600 MeV strength in the center of the nucleus, which leads to compressed
nuclear matter of ten times nuclear matter density, has been proposed
 \cite{akainew}, but a
thorough critical discussion on this work is made in \cite{toki,hyper}.
Other phenomenological potentials which are more moderately attractive (with a potential depth $\sim$ 100-200 MeV at $\rho_0$) and could in principle accommodate deeply bound states have been discussed in \cite{gal1,gal2,gal3,gal4,gal}. In these latter works  a relativistic mean field
approach is followed, introducing $\sigma$ and $\omega$ fields which couple to kaons and
nucleons to obtain the $\bar{K}$ nucleus optical potential. The same approach, but taking care of 
the momentum dependence of the potential and leading to a weaker attraction, is followed in \cite{zhong1,zhong2,zhong3}.
The philosophy in such approaches is that these fields are
average fields, not necessarily the same ones describing the $\bar{K}N$ interaction,
and the parameters of the mean field potential are fixed by fitting the data on kaonic
atoms.  Then the resulting potential is used to study the possible deeply
bound states of kaons in nuclei.  It is a fact that the observed kaonic atoms
feel the surface of the nucleus and provide information about the $K^-$ nuclear
interaction at low densities.  Any phenomenological potential fitting the atomic data has to make assumptions on
the density dependence in order to extrapolate it to higher densities and,
hence, the predictions at high densities are a consequence of the 
density shape enforced to the potential. The ambiguities of the potential at these
higher densities remains, in spite of the fact that a best fit to the atomic data
is made, and a good quality $\chi$-square is obtained as in \cite{gal}.  

    Even though the $\sigma K$ coupling in the relativistic mean field approach is
 phenomenological in nature, it is nevertheless
 interesting to see what does one obtain for the $\sigma K$ coupling at the
 elementary level.  This is possible within the chiral unitary approach,
 following similar lines to what was done in \cite{sigmapot} to get the
 contribution of $\sigma$ exchange to the $NN$ potential, or to the $\Lambda ~N$
 and $\Lambda ~ \Lambda$ in \cite{sasaki}. This is the purpose of the present
 paper. While addressing this problem, we  realize that we generate a genuine term
 that one should also take into account in the study of the $\bar{K}N$
 elementary interaction, and which is not done in the studies carried out
 with the lowest order chiral Lagrangian 
 \cite{angels,Oller:2000fj,Oset:2001cn,jido,Garcia-Recio:2002td,Garcia-Recio:2003ks,Hyodo:2002pk}.
 Interestingly, we will find that this contribution is zero at $\bar{K}N$
 threshold for $\bar{K}$ forward angles, and quite small in any case. Using this term in the construction of the
 optical potential at the mean field level for kaon interaction in infinite
 nuclear matter, which requires the forward amplitude, one obtains a null contribution. This is a consequence 
 of the fact that the leading order chiral Lagrangian gives an $I=0$ $\pi\pi\rightarrow K\bar{K}$ amplitude
 proportional to the squared momentum transfer, $q^2$. This result was already found
 in \cite{meissner} although there only the direct coupling of the kaon to the pion cloud was used, through the 
 $K\pi\rightarrow K\pi$ amplitude, while here we consider other steps involving the $\pi\pi$ interaction to make a 
 connection with the phenomenological ``$\sigma$'' exchange. 
 
     We shall also discuss the contribution of such a term in the case of kaon
 pair creation, since due to the details of chiral dynamics involved, the
 contribution is finite.

     The paper proceeds as follows: in next section we present results for the $\sigma$ coupling to $K\bar{K}$ at the
$\sigma$ pole. In section 3 and 4 we study the effects of the ``$\sigma$'' exchange in the $t$-channel for the $\bar{K} N \rightarrow \bar{K} N$ interaction. In section 5 we compare these results with the Weinberg-Tomozawa interaction and in section 6 we draw conclusions. 

\section{Coupling of $\sigma$ to $K\bar{K}$ in the $s$-channel}
The $\sigma$ meson appears in chiral unitary approaches as a dynamically
generated resonance in the interaction of $\pi\pi$ and $K\bar{K}$ coupled
channels in $I=0$ and $L=0$ \cite{npa,kaiser,markushin}. It appears as a pole
in the $t$-matrix of the $\pi\pi$ scattering in the complex plane, from where one can deduce the
mass and the width \cite{npa,pelaezprd,nsd}. These values are around 450 MeV
and 400 MeV respectively, and are in agreement with alternative approaches
based on the Roy equations \cite{colangelo}. The inspection of the poles not
only provides the mass and the width of the resonance, but also the coupling of
the resonance to the different channels by evaluating the residues of the
different amplitudes at the pole. This analysis has already been done in \cite{nsd},
where one obtains for the moduli of the couplings

\begin{equation}
\frac{|\xi^\sigma_{K\bar{K}}|}{|\xi^\sigma_{\pi\pi}|}=0.254;\quad|\xi^\sigma_{\pi\pi}|=4.26
\,GeV,\quad|\xi^\sigma_{K\bar{K}}|=1.08\, GeV .\label{gsigmaK}
 \end{equation} 
when a mixture with a possible preexisting scalar state around $1 GeV$ of mass is allowed and

\begin{equation}
\frac{|\xi^\sigma_{K\bar{K}}|}{|\xi^\sigma_{\pi\pi}|}=0.301;\quad|\xi^\sigma_{\pi\pi}|=4.21
\,GeV,\quad|\xi^\sigma_{K\bar{K}}|=1.27\, GeV 
 \end{equation} 
when the scalar mesons below $1.2$ $GeV$ are generated dynamically.

In \cite{Daniel}, where in addition scalar mesons with open and hidden charm are investigated, the couplings of the $\sigma$ have been reevaluated. Their complex value (and moduli in brackets) are listed below

\begin{equation}
\frac{\xi^\sigma_{K\bar{K}}}{\xi^\sigma_{\pi\pi}}=0.298-i0.046;\quad\xi^\sigma_{\pi\pi}=2.94-i3.04\,\, (4.23)
\,GeV,\quad\xi^\sigma_{K\bar{K}}=0.74-i1.041\,\, (1.28)\, GeV 
 \end{equation} 

These couplings have been calculated at the pole of the
$\sigma$, hence, $\xi^\sigma_{K\bar{K}}$
provides the coupling of the $\sigma$ to a virtual $K\bar{K}$ state with total
energy of $\sim$ 450 MeV at rest. Although this situation is not the one we will
encounter in the study of the $\bar{K}N\longrightarrow\bar{K}N$ interaction in
the next section, it is interesting to compare the $|\xi^\sigma_{K\bar{K}}|$ coupling
with those used in \cite{gal1,muto,amigo1,amigo2}. In \cite{gal1} the $\sigma K$ coupling
is defined by means of the Lagrangian
\begin{eqnarray}
\mathcal{L}_k = \partial_{\mu} \bar{\psi} \partial^{\mu} \psi - m_k^2 \bar{\psi} \psi - g_{\sigma K} m_k \bar{\psi} \psi \sigma
- i g_{\omega K} (\bar{\psi} \partial_{\mu} \psi \omega^{\mu} -  \psi \partial_{\mu} \bar{\psi} \omega^{\mu} ) + (g_{\omega K} \omega_{\mu})^2
\bar{\psi} \psi 
\end{eqnarray}
which describes the interaction of the antikaon field ($\bar{\psi}$) with the scalar ($\sigma$) and vector ($\omega$) isoscalar fields. We should compare $g_{\sigma K}m_K$ with $|\xi^\sigma_{K\bar{K}}|$. However, $g_{\sigma K}$ is given in     
 \cite{gal1} in terms of $\alpha_\sigma$ as
\begin{equation}
\alpha_\sigma=\frac{g_{\sigma K}}{g_{\sigma N}}.
\end{equation}
We take a standard value of $g_{\sigma N}$ from \cite{holinde}
\begin{equation}
\frac{{g_{\sigma N}}^2}{4\pi}=5.69
\end{equation}
giving  $g_{\sigma N}=8.45$. Hence, in the chiral unitary framework for the 
energy of the $\sigma$ one finds
\begin{equation}
\alpha_\sigma=\frac{|\xi^\sigma_{K\bar{K}}|}{m_K g_{\sigma N}}=0.260-0.307
\end{equation}

This result is curiously close to the one obtained by quark models, i.e., $1/3$, as quoted
in \cite{gal1}. It is also close to the value of reference $\sim
0.2-0.3$ used in \cite{gal1} and modified in various ways in subsequent papers
 \cite{gal2,gal3,gal4}.

In other works $g_{\sigma K}$ is connected to the
$\bar{K}N$ $\Sigma$-term \cite{muto,amigo1,amigo2} as

\begin{equation}
\frac{g_{\sigma N}g_{\sigma K}}{{m_\sigma}^2}=\frac{\Sigma_{\bar{K}N}}{2m_K f^2}
\end{equation}
and the value of $\Sigma_{\bar{K}N}$ is taken from some lattice calculation \cite{qcd-lat} ranging from
$\Sigma_{\bar{K}N}\sim290-450$ MeV, which gives $g_{\sigma K}\sim0.85-1.32$. In \cite{muto2} two values have been used $\Sigma_{\bar{K}N}=$ $305$, $207$ MeV. The values of $g_{\sigma K}\sim0.85-1.32$ are smaller than those used in \cite{gal1}, but they are further increased to $g_{\sigma K}\sim 2.21-2.52$ in order to obtain $\bar{K}A$ potentials of the order of $120$-$130$ MeV at $\rho=\rho_0$.

The discussion above clarifies the meaning of the $g_{\sigma K}$ couplings used in the literature. However, the information provided in eq. (\ref{gsigmaK}) for the $g_{\sigma K}$ coupling from the microscopical chiral unitary theory is not what one should use in $\bar{K}N$ scattering since, as mentioned above, this corresponds to an energy of about $450$ MeV in the $\bar{K}K$ production channel. This is the $t$-channel in $\bar{K}N\rightarrow\bar{K}N$ scattering and the equivalent $K\bar{K}$ energy, the Mandelstam variable $t$ in $\bar{K}N\rightarrow\bar{K}N$, is zero for forward $\bar{K}$ scattering. We focus the discussion for this case in the next section.

\section{$\bar{K}N\rightarrow\bar{K}N$ with $\sigma$ exchange}
\begin{figure}
\begin{center}
\resizebox{0.5\textwidth}{!}{\includegraphics{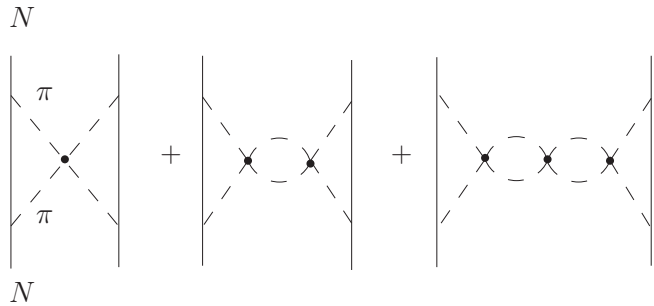}}
\caption{Diagrammatic representation of the ``$\sigma$'' exchange in the NN interaction.}\label{sigmaex}
\end{center}
\end{figure}

In \cite{sigmapot} the chiral unitary approach was used to describe microscopically the ``$\sigma$'' exchange in $NN$ collisions. Since the ``$\sigma$" is a dynamically generated resonance from the interaction of $\pi\pi$, $K\bar{K}$ (largely $\pi\pi$), the microscopic picture for $\sigma$ exchange is given in fig.  \ref{sigmaex}.

\begin{figure}
\begin{center}
\resizebox{0.5\textwidth}{!}{\includegraphics{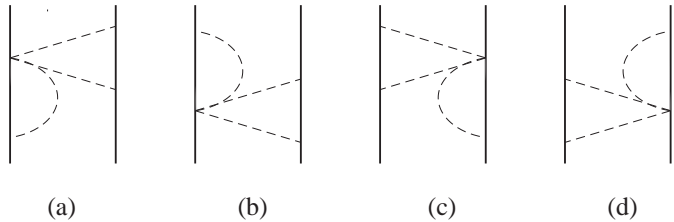}}
\caption{The processes with a three meson vertex at a baryon line for $pp\rightarrow pp$.}\label{cancel}
\end{center}
\end{figure}

\begin{figure}
\begin{center}
\resizebox{0.6\textwidth}{!}{\includegraphics{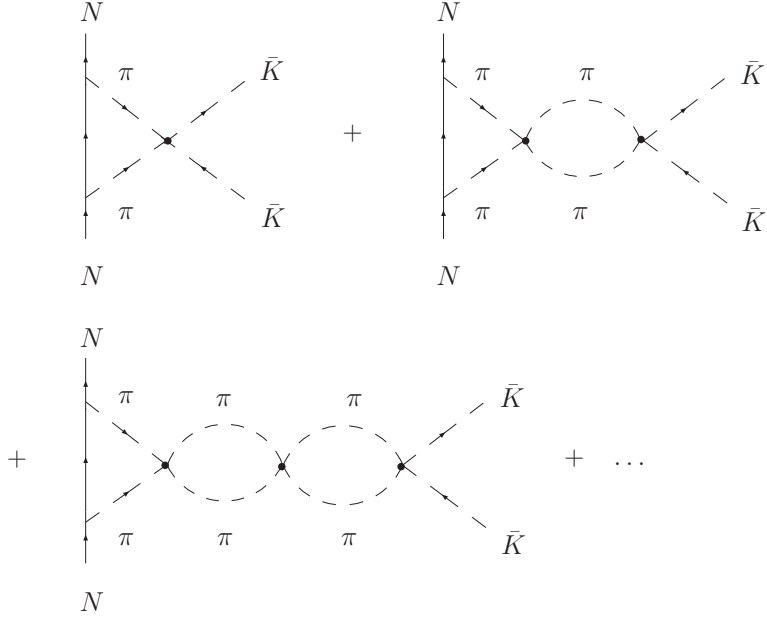}}
\caption{Diagrammatic representantion of the $\bar{K}N$ interaction via ``$\sigma$'' exchange.}\label{coupl}
\end{center}
\end{figure}

In the present context it is essential to recall that even if the $\pi\pi\rightarrow\pi\pi$ amplitude in fig.  \ref{sigmaex} appears in principle off-shell, an exact cancellation was obtained of the off-shell contribution of pions attached to the nucleons with the diagrams of the type shown in fig.  \ref{cancel}.
For the pions inside $\pi\pi$ loops the contribution of the off-shell pions in the $\pi\pi\rightarrow\pi\pi$ vertices was shown to get absorbed into the couplings and masses of the theory by renormalizing them \cite{npa}. As a consequence of all this, one has to evaluate only the diagrams of fig.  \ref{sigmaex} by using the on-shell $\pi\pi\rightarrow\pi\pi$ vertices provided by the lowest order chiral Lagrangian. This said, it is straightforward to draw the diagrams contributing to the coupling of the kaon to the nucleon via ``$\sigma$'' exchange (see fig.  \ref{coupl}).

\begin{figure}
\begin{center}
\resizebox{0.7\textwidth}{!}{\includegraphics{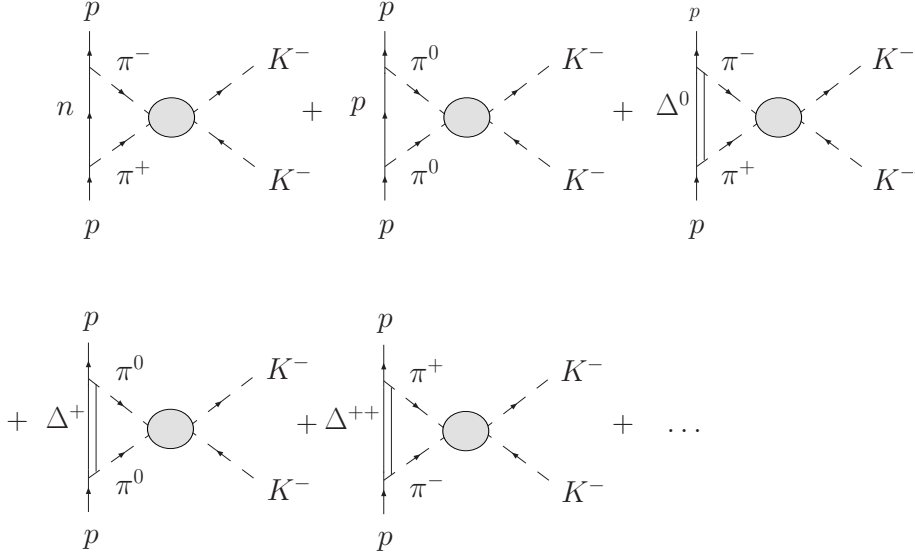}}
\caption{Diagrams contributing to the $K^- p \rightarrow K^- p$ process through ``$\sigma$'' exchange.}\label{pK}
\end{center}
\end{figure}

The diagrams for $K^- p \rightarrow K^- p$ can be drawn in detail as shown in fig.  \ref{pK},
where the circle denotes the full $\pi\pi\rightarrow K\bar{K}$ $t$-matrix, including all the bubble iterations implicit in the Bethe-Salpeter equation used to evaluate the amplitude in \cite{npa}.

Using the unitary normalization
\begin{eqnarray}
\mid\pi\pi,I=0\rangle&=&-\frac{1}{\sqrt{6}}\Big[\mid\pi^0\pi^0\rangle+\mid\pi^+\pi^-\rangle +\mid\pi^-\pi^+\rangle\Big],\nonumber\\
\mid K\bar{K}, I=0\rangle &=&-\frac{1}{\sqrt{2}}\Big[\mid K^+ K^-\rangle + \mid K^0\bar{K}^0\rangle\Big],
\end{eqnarray}
we get
\begin{equation}
t_{K^- p \rightarrow  K^- p}=\tilde{V}(q)\sqrt{3}\,t^{I=0}_{\pi\pi\rightarrow K\bar{K}} (q), \label{tpK}
\end{equation}
where $\tilde{V}(q)=\tilde{V}_N(q)+\tilde{V}_\Delta(q)$. $\tilde{V}_N(q)$, $\tilde{V}_\Delta(q)$ are the vertex functions (triangle diagrams in fig.  \ref{pK}) corresponding to $N$ or $\Delta$ intermediate states, respectively, which are given in \cite{toki} by
\begin{eqnarray}
\widetilde{V}_N(q)&=&{\int}\frac{d^3p}{(2\pi)^3}\left(\frac{D+F}{2f}\right)^2
(\vec{p}^{\ 2}+\vec{p}\cdot\vec{q})\frac{M}{E}\ \frac{1}{2}\ \frac{1}{\omega}
\ \frac{1}{\omega^{\,\prime}}\ \frac{1}{\omega+\omega^{\,\prime}}
\nonumber \\
&&\times \frac{1}{E+\omega-M}\ \frac{1}{E+\omega^{\,\prime}-M}
[\omega+\omega^{\,\prime}+E-M]\label{Vtilde}
\end{eqnarray}
with
\begin{displaymath}
E=E(\vec{p});\ \omega=\sqrt{{m_{\pi}}^2+\vec{p}^{\ 2}};\ 
\omega^{\,\prime}=\sqrt{{m_{\pi}}^2+(\vec{p}+\vec{q})^{2}}
\end{displaymath}
and
\begin{eqnarray}
\widetilde{V}_{\Delta}(q)&=&\frac{4}{9}{\left(\frac{f^*_{{\pi}N{\Delta}}}
{f_{{\pi}NN}}\right)}^2
\int\frac{d^3p}{(2\pi)^3}\left(\frac{D+F}{2f}\right)^2
(\vec{p}^{\ 2}+\vec{p}\cdot\vec{q})\frac{M_\Delta}{E_\Delta}\frac{1}{2}\frac{1}{\omega}\frac{1}{\omega^{\,\prime}}
\nonumber \\
&&\times\frac{1}{\omega+\omega^{\,\prime}}\ 
\frac{1}{E_\Delta+\omega-M}\ \frac{1}{E_\Delta+\omega^{\,\prime}-M}
[\omega+\omega^{\,\prime}+E_\Delta-M],\nonumber
\end{eqnarray}
where $q=p_{K^-_f} -p_{K^-_i}$ is the four momenta transfer, $D=0.8$, $F = 0.46$, $M_\Delta$ the $\Delta$ mass and $E_\Delta=({M_\Delta}^2+\vec{p}^{\ 2})^{1/2}$. We take the empirical value for the ratio of the
couplings $f^*_{\pi N\Delta}$ to $f_{\pi NN}$ as 2.12.

The transition amplitude of eq. (\ref{tpK}) is given by
\begin{equation}\label{t_iso_eq0}
t^{I=0}_{\pi\pi\rightarrow K\bar{K}}(q)=\frac{V^{I=0}_{\pi\pi\rightarrow K\bar{K}}(q)}{1+G(q^2)\frac{\displaystyle 1}{\displaystyle f^2}\Big(q^2-\frac{\displaystyle m^2_\pi}{\displaystyle 2}\Big)}\label{t0},
\end{equation}
which has been obtained : 1) by writing only the $\pi\pi\rightarrow\pi\pi$ scattering as the intermediate 
process (lead by the Adler amplitude $-\frac{1}{f^2}\Big(q^2-\frac{m^2_\pi}{2}\Big)$), complying with the finding 
in  \cite{npa} that the $K\bar{K}$ channel plays basically no role in building the ``$\sigma$'', and 2)  by
factorizing the last vertex, i.e, the $\pi\pi\rightarrow K\bar{K}$ transition potential
(see fig.   \ref{coupl}). The function $G(q^2)$, the loop function of two pion propagators, is given by
\begin{equation}
G(q^2)=\frac{1}{(4\pi)^2}\Big[-1+ln\frac{m^2_\pi}{\mu^2}+\beta\, ln\frac{\beta+1}{\beta-1}\Big],
\end{equation}
with $\mu$ the regularization mass, which was found to be $\mu=1.2q_{max}=1.1$ GeV \cite{pelaezprd} for the value of the cutoff, $q_{max}$, needed for a good fit to the $\pi\pi$ data and
\begin{equation}
\beta=\sqrt{1-\frac{4m^2_\pi}{q^2}}. 
\end{equation}
Let us now consider the transition potential, for the $\pi \pi \rightarrow K \bar{K}$ scattering, obtained in \cite{npa} which gives 
\begin{equation}
V^{I=0}_{\pi\pi\rightarrow K\bar{K}}(q)=-\frac{1}{3\sqrt{12}f^2}\Big(\frac{9}{2}q^2+3m^2_K+3m^2_\pi-\frac{3}{2}\sum_i p^2_i\Big).
\end{equation}
Recalling that this amplitude is to be used on-shell 
in our approach, which means taking $p_i^2 = m_i^2$ in the potential, we find
\begin{equation}\label{v_iso_0}
 V_{\pi \pi \rightarrow K \bar{K}}^{I=0} = -\frac{\sqrt{3}}{4 f^2}  q^2, 
\end{equation}
which vanishes for $K^- p$ scattering in the forward direction. The consequence is, thus, that the
$\sigma$ exchange for $K^-\, p\, \rightarrow \, K^-\, p$ scattering in the forward direction is exactly zero. We will come back to the non-forward case.

Now we discuss the case of the $K^-$-nucleus interaction in the mean field approach. A simple impulse approximation, disregarding other sources of interaction and $\bar{K}N$  rescattering effects, would give a nuclear potential such that 
\begin{equation}
V_{opt}^{``\sigma{\textrm {\scriptsize ''}}}  \rho = t_{K^-\, p\, \rightarrow \, K^-\, p} (\theta = 0) \rho = 0 
\end{equation}
with  $t_{K^-\, p\, \rightarrow \, K^-\, p}$ given by eq. (\ref{tpK}).

So far we have relied upon the exchange of pions which generate the ``$\sigma$'' through the pion interaction. We could hope that if we allowed kaons to be exchanged instead of pions, hence, building the $f_0$(980) resonance, we could get some finite contribution for $V_{opt}^{``\sigma{\textrm {\scriptsize ''}}}$, but once again we have from Eqs.(18) of \cite{npa}
\begin{equation}
V_{K \bar{K}, K \bar{K}}^{I = 0} = -\frac{1}{4 f^2} ( 3 q^2 - \sum p_i^2 + 4 m_K^2 )\,\,
\stackrel{\begin{array}{c}{\textrm{\small on-shell}}\\ \end{array}}{\Longrightarrow} -\frac{3}{4 f^2} q^2,
\end{equation}
which also vanishes identically for forward kaons, providing no contribution to the $K^-$  optical potential in nuclear matter in the impulse approximation.

The microscopic picture has, thus, given  no ``$\sigma$'' contribution to the $K^-$ nucleus optical potential. This situation is quite different from the one encountered in the NN interaction. There the $\sigma$ exchange provides a finite  contribution to the NN interaction, and in fact is quite important to account for the binding of nuclei. It is interesting to compare the $\sigma$ exchange obtained for the $NN$ interaction with the microscopic picture with the one obtained from the conventional $\sigma$ exchange. In the latter case, in momentum space one has 
\begin{equation}
V_{con}^{\sigma} (q) = g^2_{\sigma N} \frac{1}{(q^0)^2 - \vec{q}\,^2 - m_{\sigma}^2}, \,\,\,\,\, (q^0 = 0),\label{vcon}
\end{equation}
while in the microscopic case one obtains \cite{sigmapot}
\begin{eqnarray}
V_{mic}^{\sigma} (q) = [ \tilde{V} (q) ]^2 \frac{6}{f^2} \frac{\Big(\vec{q}\,^2  + \frac{\displaystyle m_{\pi}^2}{\displaystyle 2}\Big)}{ 1 - G (-\vec{q}\,^2)\frac{\displaystyle 1}{\displaystyle f^2}
\Big(\vec{q}\,^2 + \frac{\displaystyle m_{\pi}^2}{\displaystyle 2}\Big)}, \,\,(q^0 = 0, \,\,q^2 = -\vec{q}\, ^2)\label{vmicro}.
\end{eqnarray}
It is possible to establish the equivalence of the denominators in both cases since they both generate the $\sigma$ pole at the $\sigma$ mass. But then the microscopic picture contains extra $q$ dependence from the vertex functions and the numerator. In coordinate space this potential leads to an attraction at long distances compatible with the one found in    
\cite{holinde} but at short distances it produces a repulsion \cite{sigmapot}. This change of sign in the ``$\sigma$'' exchange has been tested in \cite{rijken} in the study of the hyperon nucleon interaction leading to a better fit to data. 

The differences between these approaches, $\sigma$ exchange, and our theoretical procedure, that we should rather call isoscalar correlated two pion exchange to clearly distinguish it from ordinary $\sigma$ exchange, can be traced back to essential features of the $\pi\pi$ interaction. The $\sigma$ pole indeed appears in the chiral approaches \cite{npa,kaiser,markushin,pelaezprd,nsd,pelaez2006} and it also shows up in the $\pi\pi$ scattering matrix resulting from the Roy equations analysis \cite{colangelo,caprini,colangelo2}. In all these pictures one can see that the $\pi\pi$ amplitude at low energies bears more information than the one that would result from the consideration of the $\sigma$ pole alone. This is the fundamental reason of the differences found between a simplified $\sigma$ exchange and our correlated isoscalar two pion exchange.

It should also be mentioned that the features of the correlated two pion exchange $NN$ interaction, with a weak attraction at large distances and repulsion at short distances, is tied to the way we regularize the triangle loop function of the vertex $\tilde{V}(q)$, where a cut off of around $1 GeV$ in the modulus of the three momentum is used. These features are somewhat different than those obtained in \cite{Kaiser:1998wa}, where a different renormalization is done and the $\pi\pi$ multiple scattering is not considered. The cut off used by us is of natural size but its ultimate justification lies in the fact that with this cut off one obtains  a good agreement with the empirical ``$\sigma$'' exchange potential at long distances of \cite{holinde}, and also to the fact that adding to the obtained $NN$ potential the uncorrelated two pion exchange and an empirical $\omega$-exchange potential one finds a good agreement \cite{palomar} with the empirical scalar isoscalar Argonne-$14$, -$18$ potentials \cite{Wiringa,Wiringa2}.

Coming back to the present case, the $\sigma$ exchange  $\bar{K}N$ potential has only one vertex $\tilde{V} (q)$ and the numerator is also different than in the case of the NN interaction, since the numerator of eq. (\ref{vmicro}) does not vanish at $q^2$ = 0 while the one in eq. (\ref{t0}) vanishes in that limit. The relativistic mean field approach derives the  $\sigma K$ potential by solving the equations of motion with some effective Lagrangians in the mean field approximation. The  $\sigma K$ potential is proportional to $\sigma (r) g_{\sigma K}$, with $\sigma (r)$ the $\sigma$ mean field. This field bears some memory of the $\sigma$ propagator of our approach in coordinate space. The coupling of $\sigma (r)$ to $K$ is factorized out in the mean field potential and in \cite{gal1} $g_{\sigma K}$ is varied in some range, multiplying the free value of reference by some factor. The case here is that the $\sigma K$ coupling is zero at the elementary level and by multiplying it by $\sigma (r)$ and any other factor it will remain zero.

In the present case, unlike the case of the NN interaction, the effective elementary $g_{\sigma K}$ is zero and cannot serve as a reference of a size to be moderately changed when going to the nuclear case. 

We can rephrase some of the former ideas as follows.
In the relativistic mean field approach for nucleons in nuclear matter, the effective $\sigma$ exchange is not exactly the same as in the NN interaction, but bears some memory of it and one hopes that some microscopical derivation could establish the link. In the present case there 
is no $\sigma$ exchange contribution at the elementary level in the $K^-\, N\, \rightarrow \, K^-\, N$ amplitude for forward angles,  which leaves us wondering what is the meaning of the $\sigma$ exchange at the mean field nuclear level. One would have to go through  the iterations of the Weinberg-Tomozawa dominant $K^-\, N\, \rightarrow \, K^-\, N$ potential through intermediate $\bar{K} N$ and $\pi \Sigma$ states to generate the $K^-\, N\, \rightarrow \, K^-\, N$ $t$-matrix as a first step to construct the $K^-$ nucleus optical potential, but these diagrams do not lead to a structure resembling the ``$\sigma$'' exchange in the t-channel. 

\section{Additional channels and higher order contributions.}
So far our discussion relies upon the use of the lowest order Lagrangian and the $\pi\pi$ and $K\bar{K}$ channels. We must take into account that if we include the $\eta\eta$ channel, the coupling of $\eta\eta$ to $K\bar{K}$ is not null at forward angles. In order to estimate its contribution it suffices to evaluate the first diagram of fig.  \ref{coupl}, replacing the two pions with two $\eta$'s. 

The contribution of this term to eq. (\ref{tpK}) would be given by
\begin{equation}
t_{pK^- \rightarrow p K^-}^{(\eta)}=\tilde{V}(q)t_{\eta\eta\rightarrow K\bar{K}(q)}\label{teta}
\end{equation}
with $\tilde{V}(q)$ obtained from eq. (\ref{Vtilde}) substituting 
\begin{equation}
\frac{D+F}{2f} \Rightarrow \frac{1}{\sqrt{3}}\frac{3F -D}{2f},
\end{equation}
\begin{equation}
 \omega \Rightarrow \sqrt{ m_\eta^2 + \vec{p}^2}; \quad \quad \omega^{\,\prime} \Rightarrow \sqrt{ m_\eta^2 + 
 (\vec{p}+\vec{q})^2}.
\end{equation}
The amplitude $t_{\eta\eta \rightarrow K \bar{K}}$ is given at the lowest order by
\begin{equation}
t_{\eta\eta \rightarrow K \bar{K}} = \frac{6 m_\eta^2 + 2 m_\pi^2 -9 q^2}{12 f^2}. \label{Veta}
\end{equation}

The evaluation of eq. (\ref{teta})  gives us 
\begin{equation}
t_{pK^- \rightarrow p K^-} = 1.4 \times 10^{-4} MeV^{-1}\quad(q^2 = 0).
\end{equation}

This should be compared with the  isoscalar $t_{K^- N  \rightarrow K^- N}$ amplitude at the $\bar{K}N$ threshold \cite{angels,borasoy} of the order of 
\begin{equation}
t_{K^- N  \rightarrow K^- N}^{isos} \simeq 0.025 MeV^{-1}
\end{equation}

or the equivalent $t$-matrix assuming the $K^-$ selfenergy at normal nuclear matter density
\begin{equation}
\Pi = 2 m_K V_{opt} = 2 m_K (-50 MeV) \simeq t^{eff} \rho^0
\end{equation}
giving 
\begin{equation}
t^{eff} \simeq -0.038 MeV^{-1}.\label{eff}
\end{equation}
The change of sign from the free $t$-matrix to the effective one in the medium is a consequence of the Pauli blocking and selfconsistency  in the many body calculation \cite{lutz,angelsself, schaffner}. Hence, the $\eta$ channel contribution is of the order of 1$\%$ and, thus, negligible. Other contributions involving previous $\pi\pi$ scatterings in the series of fig.  \ref{coupl} give even smaller contributions.

Next we turn to higher order contribution in the $\pi\pi \rightarrow K\bar{K}$ amplitude. This contribution does not vanish at the threshold. Results for the $\pi\pi \rightarrow K\bar{K}$ amplitude can be found in the appendix of \cite{pelaezprd}.

Using the second order chiral Lagrangian of Gasser and Leutwyler \cite{glnp}, we find for the $I=0$ case \cite{pelaezprd}

\begin{equation}
T^{(0)} = \frac{\sqrt{3}}{2} [ T^{3/2} (u,s,t) + T^{3/2} (t,s,u)]
\end{equation}
with

\begin{eqnarray}
T_4^{(3/2)} (s,t,u) &=&-\frac{2}{\fpi\fk} \left\{ (4L_1+L_3)(t-2\mpi)(t-2\mk)
+2L_2(\sq-s)^2\right.\nn\\
&+&\left.(2L_2+L_3)(\sq-u)^2+4L_4\left[(\sq)t-4\mpi\mk\right]\right.\nn\\
&+&\left.L_5\left[(\sq)(\sq-s)-4\mpi\mk\right]+8\mpi\mk(2L_6+L_8)\right\}\nn.
\end{eqnarray}

In the forward direction of $\pi\pi \rightarrow K \bar{K}$, ($s \equiv 0$), we obtain

\begin{eqnarray}
T_4^{(0, forward)} (\pi \pi \rightarrow K \bar{K})  = -\frac{\sqrt{3}}{f^4} 8 m_\pi^2 m_K^2 \{ 4 L_1 + L_3 - 4 L_4 - L_5 + 4 L_6 + 2 L_8 \}
\end{eqnarray}
and using the numerical values of \cite{glnp} we find for the combination of $L_i$ coefficients within the curly brackets

\begin{equation}
\left\{ L_i \right\} = [ -1.1 \pm  1.7 ] \times 10^{-3}.
\end{equation}

As we can see, the result is compatible with zero. Yet, it is important to see the order of magnitude of this contribution.
Coming back to eq. (\ref{tpK}) and substituting there the new $T_4$  contribution to the $\pi \pi \rightarrow K \bar{K}$ amplitude, we find 
\begin{equation}
t_{4}(K^- p\rightarrow K^- p,q^2=0) \simeq 0.0014 MeV^{-1}
\end{equation}
which is about a factor thirty smaller than the effective $t$-matrix of eq. ({\ref{eff}}) and, hence, negligible since it is far smaller than theoretical uncertainties from other sources \cite{Borasoy:2005ie}.

\section{Comparison of the ``$\sigma$'' exchange term with the Weinberg-Tomozawa term}
\begin{figure}
\begin{center}
\resizebox{0.5\textwidth}{!}{\includegraphics{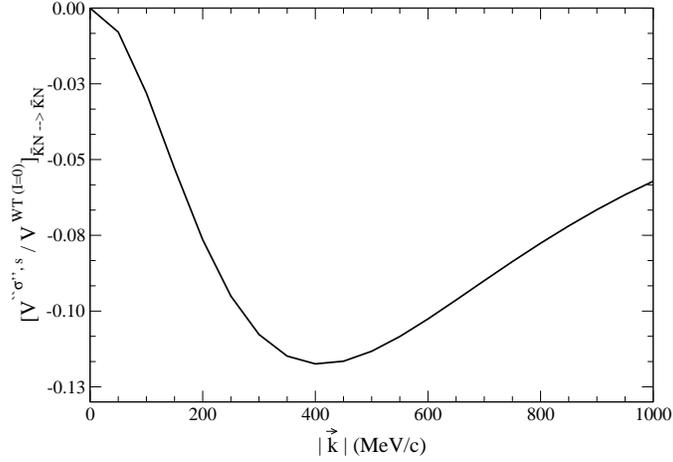}}
\caption{Comparison of the  $K^- N$ potential obtained from ``$\sigma$'' exchange with the Weinberg-Tomozawa potential.}\label{ratio}
\end{center}
\end{figure}

We now compare eq. (\ref{tpK}) with the I = 0 Weinberg-Tomozawa (WT) term for $K^-\, N\, \rightarrow \, K^-\, N$ potential in S-wave
\begin{equation}
V^{WT (I=0)}_{K^-\, N\, \rightarrow \, K^-\, N} = - \frac{3}{4 f^2} k^0 
\end{equation}
with $k^0=\sqrt{\vec{k}^2+m^2_K}$, the kaon energy in the $C.M.$ frame. On the other hand, eq. (\ref{tpK}), together with Eqs. (\ref{t_iso_eq0}) and (\ref{v_iso_0}), can be written as
\begin{equation}
V^{``\sigma{\textrm {\scriptsize ''}}}_{K^-\, N\, \rightarrow \, K^-\, N} = - \frac{\frac{3}{4f^2} q^2 \tilde{V}(q)}{ 1 + G(q^2) \frac{\displaystyle 1}{\displaystyle f^2} \Big(q^2 - \frac{\displaystyle m^2_{\pi}}{\displaystyle 2}\Big)}.\label{Vsig}
\end{equation}
The value $q^2$ in the ``$\sigma$'' exchange potential is
\begin{equation}
q^2 = (k^{\prime 0} - k^0 )^2 - (\vec{k}^{\prime} - \vec{k})^2 = - 2 \vec{k}^2 (1 - cos(\theta)),
\end{equation}
which splits this term into the s-wave and p-wave parts. At threshold this is also zero since $\vec{k} = 0$. We can project $V^{``\sigma{\textrm {\scriptsize ''}}}_{\bar{K} N \rightarrow \bar{K} N} (q)$ of eq. (\ref{Vsig}) in s-wave as
\begin{equation}
V^{``\sigma{\textrm {\scriptsize ''}}, s}_{\bar{K} N \rightarrow \bar{K} N} = \frac{1}{2} \int_{-1}^{1} dcos(\theta) V^{``\sigma{\textrm {\scriptsize ''}}}_{\bar{K} N \rightarrow \bar{K} N}.
\end{equation}
In fig.  \ref{ratio} we plot the ratio of $V^{``\sigma{\textrm {\scriptsize ''}}}_{\bar{K} N \rightarrow \bar{K} N}$ to $V^{WT (I=0)}_{\bar{K} N \rightarrow \bar{K} N}$ as a function of $|\vec{k}|$, the kaon momentum in the $\bar{K}N$ $C.M.$ frame.
As we can see, the ratio of the ``$\sigma$'' exchange contribution to the WT term is zero at the threshold and stays small in a wide range. At $|\vec{k}| \sim$ 200 MeV/c this ratio is still of the order of 8$\%$. This contribution is smaller than differences between different options of the $t$-matrix compatible with $\bar{K}N$ data, as shown in \cite{borasoy_meisner}, so there is not much to worry about, unless one is aiming at a precision higher than that provided by the actual data. This finding comes to stress once more the large dominances of the $WT$ contribution in the $\bar{K}N$ scattering, which is a basic assumption in the various chiral approaches.

\section{Conclusions}
We have evaluated the correlated isoscalar two pion exchange contribution to the $K^-\, N\, \rightarrow \, K^-\, N$ scattering within the chiral unitary approach and have found it to vanish at the threshold and at any energy in the forward direction using the lowest order of the chiral Lagrangians. This is a consequence of the fact that the leading order chiral Lagrangian gives an $I=0$ $\pi\pi\rightarrow K\bar{K}$ amplitude proportional to the squared momentum transfer, $q^2$. We also found that non vanishing forward contributions from higher order meson meson Lagrangians, or the consideration of the $\eta\eta\rightarrow K\bar{K}$ transition, led to negligible corrections.

If one makes a mapping of this elementary correlated two pion exchange to get the correlated two pion exchange contribution to the $K^-$ optical potential for $K^-$ in the nuclear matter, the potential vanishes, since it requires the elementary amplitude at forward angles. This result raises questions on the microscopical meaning of the ``$\sigma$'' exchange potential often used in the literature in relativistic mean field models of the $K^-$ nucleus interaction. On the other hand, projecting the s-wave part of the correlated isoscalar two pion exchange amplitude, we have compared it with the dominant Weinberg-Tomozawa amplitude and we find  the ratio to be quite small. It is zero at the threshold and reaches a maximum value of $<$ 12$\%$ at relatively high momenta of the kaon of the order of 400 MeV/c. This contribution is small compared with present uncertainties in the theoretical models but might become relevant in a future if one aims at having fits to increasingly more accurate data.

\section{Acknowledgments}

A.M.T. wishes to acknowledge  the support of the Ministerio de Educaci\'on y Ciencia in the program of FPU.
This work is partly supported by
DGICYT contract number FIS2006-03438,
and the Generalitat Valenciana.
This research is  part of
the EU Integrated Infrastructure Initiative  Hadron Physics Project under
contract number RII3-CT-2004-506078, and K.P.K. wishes to acknowledge direct support from it.

\end{document}